\documentclass[a4paper,11pt]{article}
\usepackage{jcappub}

\usepackage{xcolor}
\usepackage{graphicx}
\graphicspath{{./fig/}{./fig/diagram/}}
\usepackage{amsmath,amssymb,amsfonts,mathrsfs,bm}
\usepackage[%
colorlinks=true,
linkcolor=blue,
citecolor=blue,
]{hyperref}

\newcommand{\figref}{figure~\ref}

\newcommand{\dif}{\,\mathrm{d}}
\newcommand{\ee}{\mathrm{e}}

\newcommand{\meter}{\,\mathrm{m}}

\newcommand{\kpc}{\,\mathrm{kpc}}

\newcommand{\Msun}{M_{\odot}}

\newcommand{\dL}{d_{\text{L}}}
\newcommand{\dS}{d_{\text{S}}}
\newcommand{\RE}{R_{\text{E}}}
\newcommand{\REdPBH}{R_{\text{E}}^{(\text{dPBH})}}
\newcommand{\tE}{t_{\text{E}}}
\newcommand{\tEi}{t_{\text{E},i}}
\newcommand{\Rtube}{R_{\text{tube}}}
\newcommand{\Rhalo}{R_{\text{halo}}}
\newcommand{\Mhalo}{M_{\text{halo}}}
\newcommand{\MPBH}{M_{\text{PBH}}}
\newcommand{\MdPBH}{M_{\text{dPBH}}}
\newcommand{\RSchPBH}{R_{\text{Sch}}^{(\text{PBH})}}
\newcommand{\RSchSun}{R_{\text{Sch}}^{(\text{Sun})}}
\newcommand{\rS}{r_{\text{S}}}
\newcommand{\fdPBH}{f_{\text{dPBH}}}
\newcommand{\fPBH}{f_{\text{PBH}}}
\newcommand{\hatb}{\hat{b}}
\newcommand{\hatl}{\hat{l}}
\newcommand{\hatRhalo}{\hat{R}_{\text{halo}}}
\newcommand{\hatRtube}{\hat{R}_{\text{tube}}}
\newcommand{\hatrS}{\hat{r}_{\text{S}}}

\newcommand{\tot}{\text{tot}}
\newcommand{\lens}{\text{lens}}
\newcommand{\obs}{\text{obs}}
\newcommand{\DM}{\text{DM}}
\newcommand{\PBH}{\text{PBH}}
\newcommand{\NFW}{\text{NFW}}

\newcommand{\Dphi}{\Delta\phi}

\begin{document}

\title{Gravitational microlensing by dressed primordial black holes}

\author[a,b,c]{Rong-Gen Cai,}
\author[a,b,*]{Tan Chen,}
\author[a]{Shao-Jiang Wang,}
\author[d,*]{Xing-Yu Yang\note[*]{Corresponding author.}}

\affiliation[a]{CAS Key Laboratory of Theoretical Physics, Institute of Theoretical Physics, Chinese Academy of Sciences, Beijing 100190, China}
\affiliation[b]{School of Physical Sciences, University of Chinese Academy of Sciences, Beijing 100049, China}
\affiliation[c]{School of Fundamental Physics and Mathematical Sciences, Hangzhou Institute for Advanced Study, University of Chinese Academy of Sciences, Hangzhou 310024, China}
\affiliation[d]{Korea Institute for Advanced Study, Seoul 02455, Republic of Korea}

\emailAdd{cairg@itp.ac.cn}
\emailAdd{chentan@itp.ac.cn}
\emailAdd{schwang@itp.ac.cn}
\emailAdd{xingyuyang@kias.re.kr}

\abstract{%
    The accretion of dark matter around the primordial black holes (PBHs) could lead to the formation of surrounding minihalos, whose mass can be several orders of magnitude higher than the central PBH mass.
    The gravitational microlensing produced by such dressed PBHs could be quite different from that of the bare PBHs, which may significantly affect the constraints on the PBH abundance.
    In this paper, we study the gravitational microlensing produced by dressed PBHs in detail.
    We find that all the microlensing effects by dressed PBHs have asymptotic behavior depending on the minihalo size, which can be used to predict the microlensing effects by comparing the halo size with the Einstein radius.
    When the minihalo radius and the Einstein radius are comparable, the effect of the density distribution of the halo is significant to the microlensing.
    Applying the stellar microlensing by dressed PBHs to the data of the Optical Gravitational Lensing Experiment and Subaru/HSC Andromeda observations, we obtain the improved constraints on the PBH abundance.
     It shows that the existence of dark matter minihalos surrounding PBHs can strengthen the constraints on the PBH abundance from stellar microlensing by several orders, and can shift the constraints to the well-known asteroid mass window where PBHs can constitute all the dark matter.
}

\maketitle

\section{Introduction}\label{sec:intro}

The nature of dark matter is one of the most important questions of fundamental physics.
Although dark matter has not been directly detected, observational evidences such as galaxy rotation curves~\cite{1980ApJ...238..471R, Corbelli:1999af}, galaxy clusters~\cite{Allen:2011zs} and gravitational lensing~\cite{Refregier:2003ct} have implied its existence.
Many dark matter candidates were proposed, including particles beyond the standard model of particle physics and astrophysical objects~\cite{Bertone:2016nfn}.

The primordial black holes (PBHs) from the gravitational collapse of the overdense regions in the early universe~\cite{Zeldovich:1967lct, Hawking:1971ei, Carr:1974nx} can be an attractive candidate for dark matter since they may constitute all the dark matter without invoking a new set of particles.
The fraction of PBHs in the dark matter has been constrained over many mass ranges by different methods, such as black hole evaporation~\cite{Boudaud:2018hqb, DeRocco:2019fjq, Laha:2019ssq}, gravitational microlensing~\cite{EROS-2:2006ryy, Griest:2013aaa, Niikura:2017zjd, Blaineau:2022nhy, Oguri:2017ock, Niikura:2019kqi, Oguri:2022fir}, gravitational waves~\cite{LIGOScientific:2019kan, Hutsi:2020sol}, dynamical effects~\cite{Koushiappas:2017chw}, and cosmic microwave background~\cite{Poulin:2017bwe, Serpico:2020ehh}.

Gravitational microlensing is a powerful method to probe dark matter in the Milky Way (MW)~\cite{Paczynski:1985jf, MACHO:1990npi}.
It can constrain the abundance of PBHs with mass in the range of $[10^{-11},10^3]\Msun$ by detecting the corresponding microlensing event rate, and the resulting constraints indicate that PBHs in this mass range constitute a subdominant dark matter component~\cite{EROS-2:2006ryy, Niikura:2017zjd, Niikura:2019kqi, Blaineau:2022nhy}.
The PBHs, as local overdensities in the dark matter, can lead to the formation of minihalos surrounding themselves.
Such PBHs with minihalos are usually referred to as the dressed PBHs and are studied widely~\cite{Bertschinger:1985pd, Eroshenko:2016yve, Boucenna:2017ghj, Adamek:2019gns, Hertzberg:2019exb, Cai:2020fnq}.
Since the mass of the surrounding minihalo can be several orders of magnitude greater than the central PBH mass~\cite{Mack:2006gz, Ricotti:2007au}, the microlensing produced by dressed PBHs could be quite different from the bare PBHs, which may significantly affect the constraints on PBH abundance.

In this paper, we study the gravitational microlensing by dressed PBHs in detail, including the deflection angle, magnification, and the finite source size effect, which are presented in Sec.~\ref{sec:microdPBH}.
Applying the stellar microlensing by dressed PBHs to the data of Optical Gravitational Lensing Experiment (OGLE) and Subaru/HSC Andromeda observations, we obtain the improved constraints on the PBH abundance, which are presented in Sec.~\ref{sec:PBHcons}.
In Sec.~\ref{sec:microlensing} we briefly revisit the gravitational microlensing theory, and Sec.~\ref{sec:conclu} devotes to conclusion and discussion.
We set $ c = G = 1 $ throughout the paper for brevity.

\section{Gravitational microlensing}\label{sec:microlensing}

The deflection of light by the gravitational field is one of the most important predictions of general relativity and is the basis of gravitational lensing theory.
Gravitational microlensing is a particular gravitational lensing phenomenon in which the multiple images of a source are unresolved so that only the enhancement of the apparent brightness can be observed.
%The reason for the microlensing to be observed is that the magnification of the unresolved images is time-dependent due to the relative motion of the lens and the source.
In this section, we briefly revisit the gravitational microlensing theory.

\subsection{Deflection angle and imaging equation}\label{sec:deflectionangle}

The deflection angle of light has been studied for a long time.
One of the most famous results is that for a point-like lens with mass $ M $, the deflection angle of light with impact parameter $ b $ can be approximated by $ 4M/b $ under the weak field approximation.

Since the size of the lens is typically much smaller than the distances between the source, observer, and lens, it can be approximated as a thin screen with surface mass density $\Sigma (\bm{b})$, where $ \bm{b} $ represents the position of light intersecting the lens plane.
Assuming that the lens is spherically symmetric, the deflection angle $ \Dphi $ can be approximated as the sum of the deflections produced by all the mass elements under weak field approximation~\cite{Narayan:1996ba},
\begin{equation}\label{eq:deflection}
    \Dphi (b) =\frac{4 M(b)}{b},
\end{equation}
where $ M (b) $ is the mass within radius $ b $.

\begin{figure}[htbp]
    \centering
    \includegraphics[width=0.7\textwidth]{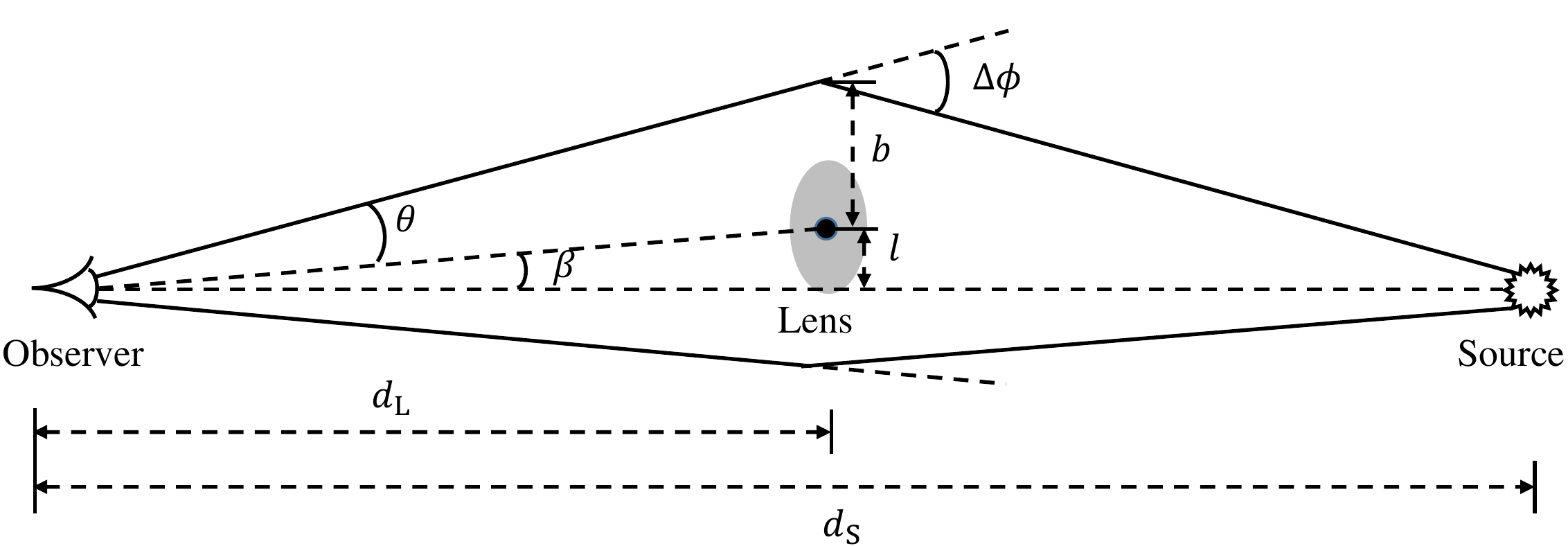}
    \caption{%
        The geometry of the lens system.
    }
    \label{fig:geoimage}
\end{figure}

The imaging equation is
\begin{equation}\label{eq:image}
    D \Dphi (b_i)=b_i \pm l,
\end{equation}
where $ D \equiv {\dL(\dS-\dL)}/{\dS} $, $\dL$ and $\dS $ are the distances from the observer to the lens and source respectively, $l$ is the separation between the lens and the line-of-sight direction, and $ \{b_i\} $ are the positions of different images.
Different signs represent different sides as shown in \figref{fig:geoimage}.
%Depending on the functional form of deflection angle $ \Dphi(b) $ given by the mass distribution of the lens, there could be multiple images for one source.

For a point-like lens with mass $ M $, there are always two images $ \{ b_1 , b_2 \} $ on either side of the line-of-sight,
\begin{equation}\label{eq:imagep}
    b_{1,2} = \frac{\mp l + \sqrt{l^2 + 4 \RE^2}}{2},
\end{equation}
where $ \RE = \sqrt{4M D}$ is the Einstein radius.

\subsection{Magnification and the finite source size effect}\label{sec:magfinite}

For a point-like source, the magnification $ \mu_i $ of one image can be calculated as~\cite{Virbhadra:1999nm}
\begin{equation}
    \mu_i =(\frac{\sin\beta \dif \beta}{\sin\theta_i \dif \theta_i})^{-1},
\end{equation}
where the angle $ \beta $ (angle $ \theta_i $) is the angular separation between the source (image) and the lens.
The total magnification is
\begin{equation}\label{eq:mag}
    \mu_{\tot}
    = \sum_{i} \left | (\frac{\sin\beta \dif \beta}{\sin\theta \dif \theta})^{-1} \right |
    \approx \sum_{i} \left | \frac{b_i \dif b_i}{l \dif l} \right |
    = \sum_{i} \frac{b_i}{l \left | \frac{\dif \Dphi}{\dif b_i} D -1 \right |}.
\end{equation}
Using eqs.~\eqref{eq:imagep} and~\eqref{eq:mag}, the magnification for the point-like source and lens can be written as
\begin{equation}\label{eq:pointmag}
    \mu_{\tot}^{\text{(p)}}=\frac{u^2+2}{u\sqrt{u^2+4}},
\end{equation}
where $ u=l/\RE $.
When $ l = \RE $, the total magnification is $ \mu_{\tot}^{\text{(p)}} = 3/\sqrt{5} \approx 1.34 $.

One can approximate the source as a point when the source size is much smaller than the Einstein radius.
However, when considering small mass lenses with the Einstein radius comparable with or smaller than the size of the source, the finite source size effect will be important.
This effect has been discussed in~\cite{1994ApJ...430..505W,Montero-Camacho:2019jte,Croon:2020wpr,Croon:2020ouk,Fujikura:2021omw}. The main idea is that one can decompose the source into point-like pieces and the image of the source will be the total of their individual images.

\begin{figure}[htbp]
    \centering
    \includegraphics[width=0.6\textwidth]{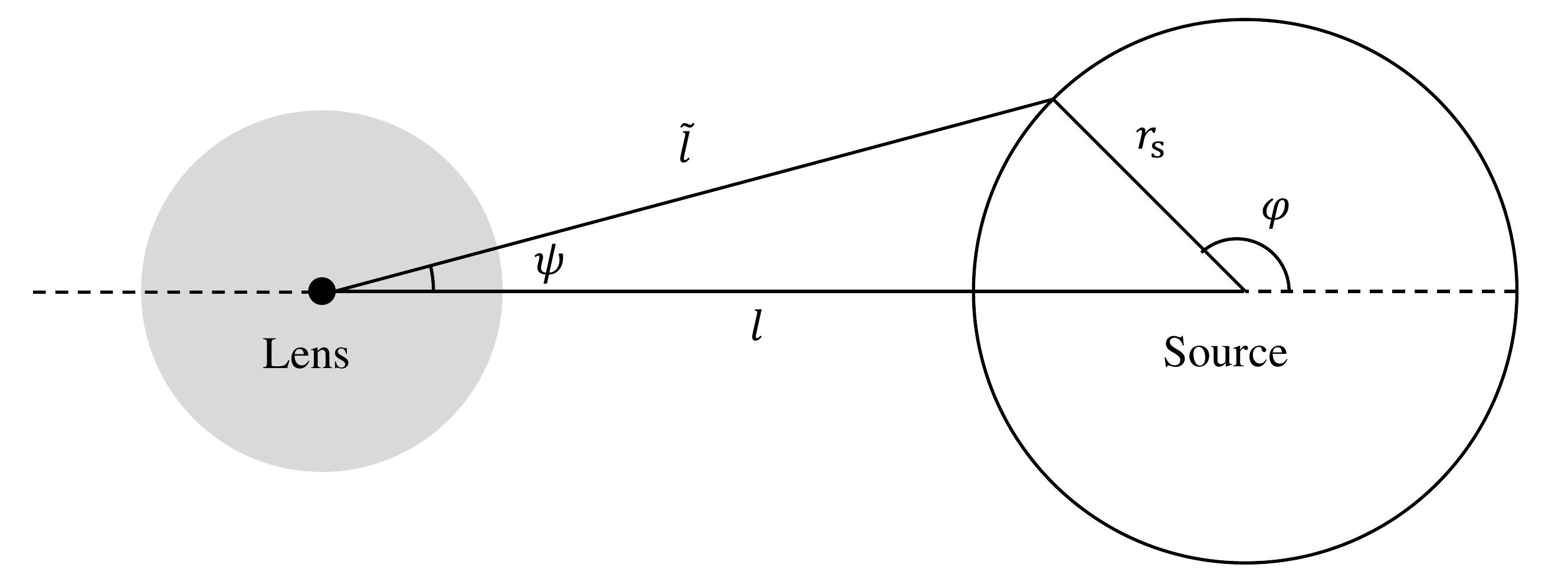}
    \caption{%
        The lens and projected source in the lens plane.
    }
    \label{fig:projectedplane}
\end{figure}

For simplicity, we assume that the source is spherically symmetric and has a uniform intensity in the lens plane.
To get the separation between the source and the lens, one needs to project the source into the lens plane as shown in \figref{fig:projectedplane}.
The distance from the lens center to a point on the edge of the projected source is
\begin{equation}
    \tilde{l}(\varphi) = \sqrt{l^2+r_{\text{S}}^2+ 2\, l \, r_{\text{S}} \cos \varphi},
\end{equation}
where $ r_{\text{S}} \equiv \frac{\dL}{\dS} R_{\text{S}} $ is the projected source radius, $ R_{\text{S}} $ is the source radius, $ l $ is the separation between the centers of the lens and projected source, and $ \varphi $ is the angular position of the point.

Then the imaging equation becomes
\begin{equation}
    D \Dphi (b_i)=b_i \pm \tilde{l}.
\end{equation}
The magnification can be calculated as the ratio of the solid angles subtended by the image to the source, which can reduce to the ratio of the image's area to the source's area in the lens plane.
Therefore the magnification of image $ b_i $ is~\cite{Fujikura:2021omw}
\begin{equation}\label{eq:finitemag}
    \mu_i = \frac{(-1)^{P_i}}{\pi r_{\text{S}} ^2} \int_{0}^{2 \pi} \dif\varphi \frac{\dif\psi}{\dif\varphi} \frac{1}{2} b_i ^2 (\varphi),
\end{equation}
with
\begin{equation}
    \tan \psi = \frac{r_{\text{S}} \sin \varphi}{l + r_{\text{S}} \cos \varphi},
\end{equation}
where $ \psi $ represents the angular position of the corresponding point on the edge of the image.
%Due to the spherical symmetry, the lens center, a point of source, and its corresponding image in the lens plane are collinear.
The relation between $ \psi $ and $ \varphi $ can be obtained by the geometric relation shown in \figref{fig:projectedplane}.
The factor $ (-1)^{P_i} $ represents the ``parity'' of the image.
According to different imaging processes, the integral in eq.~\eqref{eq:finitemag} may have the opposite sign to the real magnification, which can be corrected by this factor.
For a point-like lens, the ``parities'' of two images are opposite.
%The ``parity'' of $b_1$ is negative and the ``parity'' of $b_2$ is positive.
The total magnification is the sum of $ \{ \mu_i \} $.

\subsection{Wave optics effect}

In section~\ref{sec:deflectionangle}-\ref{sec:magfinite}, we have introduced gravitational microlensing under the geometrical optics approximation.
This approximation is reasonable in most microlensing search observations since the lens size is typically much larger than the optical wavelengths used in these observations, thus the wave optics effect can be ignored.
However, the wave optics effect needs to be considered if a PBH is light enough (smaller than $10^{-10} \Msun$) that its Schwarzschild radius is comparable with or shorter than the optical wavelengths\cite{Sugiyama:2019dgt}.
In the following, we briefly review the wave optics effect on microlensing. The detailed introduction can be seen in \cite{1992grle.book.....S, Takahashi:2003ix, Sugiyama:2019dgt}.

We first consider the wave optics effect on microlensing for a point-like source. The lensed signal amplitude can be expressed as $ \tilde{\phi}_L (f) = F(f) \tilde{\phi}(f) $, where $ f $ is the frequency of the signal, $\tilde{\phi}$ is the unlensed signal amplitude and $ F(f) $ is the amplification factor. Under the same thin-lens approximation, the amplification factor can be expressed as~\cite{1992grle.book.....S}
\begin{equation}\label{eq:waveoptics}
    F(\omega,\bm{y}) = \frac{\omega}{2 \pi i} \int \dif^2 \bm{x} e^{i \omega \phi_F (\bm{x},\bm{y})},
\end{equation}
where $ \bm{x} = \frac{\bm{b}}{\xi_0} $ is a dimensionless vector in the lens plane and $ \bm{y} = \frac{\bm{l}}{\xi_0} $ is a dimensionless vector representing the source's position in the lens plane. Vector $ \bm{b} $ is a general position in the lens plane, $ \bm{l} $ is the projected source's position in the lens plane, and $ \xi_0 $ is a characteristic length scale. The dimensionless frequency $ \omega $ is defined as
\begin{equation}
    \omega \equiv \frac{2 \pi f \dS}{\dL (\dS-\dL)} \xi_0 ^2 ,
\end{equation}
and the Fermat potential $ \phi_F $ is
\begin{equation}
    \phi_F (\bm{x},\bm{y}) = \frac{1}{2} | \bm{x} - \bm{y} |^2 - \psi_D (\bm{x}).
\end{equation}
Here we ignore an overall phase factor since we only care about the magnification of the signal.
$ \psi_D (\bm{x}) $ is the deflection potential whose derivative gives the deflection angle. $ \psi_D (\bm{x}) $ depends on the matter distribution projected on the lens plane and can be expressed as\cite{1992grle.book.....S}
\begin{equation}
    \psi_D (\bm{x}) = \frac{1}{\pi} \int \dif^2 \bm{x'} \kappa (\bm{x'}) \ln | \bm{x} - \bm{x'} |,
\end{equation}
where $ \kappa (\bm{x}) $ is the dimensionless surface mass density, which is defined as
\begin{equation}
    \kappa (\bm{x}) = \frac{\Sigma (\xi_0 \bm{x})}{\Sigma_{\mathrm{cr}}},
\end{equation}
where $ \Sigma (\bm{b}) $ is the projected surface mass density in the lens plane and $ \Sigma_{\mathrm{cr}} = \frac{\dS}{4 \pi \dL (\dS-\dL)} $ is the critical surface mass density.
Then the magnification for a point-like source is $ \mu_{\mathrm{wave}} ^{\mathrm{(p)}} (\omega,\bm{y}) =  | F(\omega,\bm{y}) |^2 $.

When the lens is light, we also need to consider the finite source size effect. The magnification for a finite-size source is \cite{Sugiyama:2019dgt}
\begin{equation}\label{eq:magfinitewave}
    \mu_{\mathrm{wave}} (\omega,\bm{y},r_\xi) = \frac{1}{\pi r_\xi^2} \int_{| \bm{r} | \leq r_\xi} \dif^2 \bm{r} \mu_{\mathrm{wave}}^{\mathrm{(p)}} (\omega,\bm{y} - \bm{r}),
\end{equation}
where $r_\xi = \frac{\rS}{\xi_0} $ is the dimensionless source radius.

%According to eq.~\eqref{eq:magfinitewave}, the finite source effect will cause the magnification of a point-like source to average out and reduce its volatility.
%Therefore the finite source size effect can counteract the wave optics effect. 
%The finite source size effect is usually dominant over the wave optics effect~\cite{Sugiyama:2019dgt, Croon:2020ouk}.
%So we do not analyze the wave optics effect in detail. We only consider the wave optics effect when we calculate the constraints for light dressed PBHs. 

%According to eq.~\eqref{eq:magfinitewave}, the finite source effect will average out the magnification of a point-like source.
%Usually the finite source size effect is dominant over the wave optics effect.

\subsection{Event rate of microlensing}\label{sec:eventrate}

\begin{figure}[htbp]
    \centering
    \includegraphics[width=0.8\textwidth]{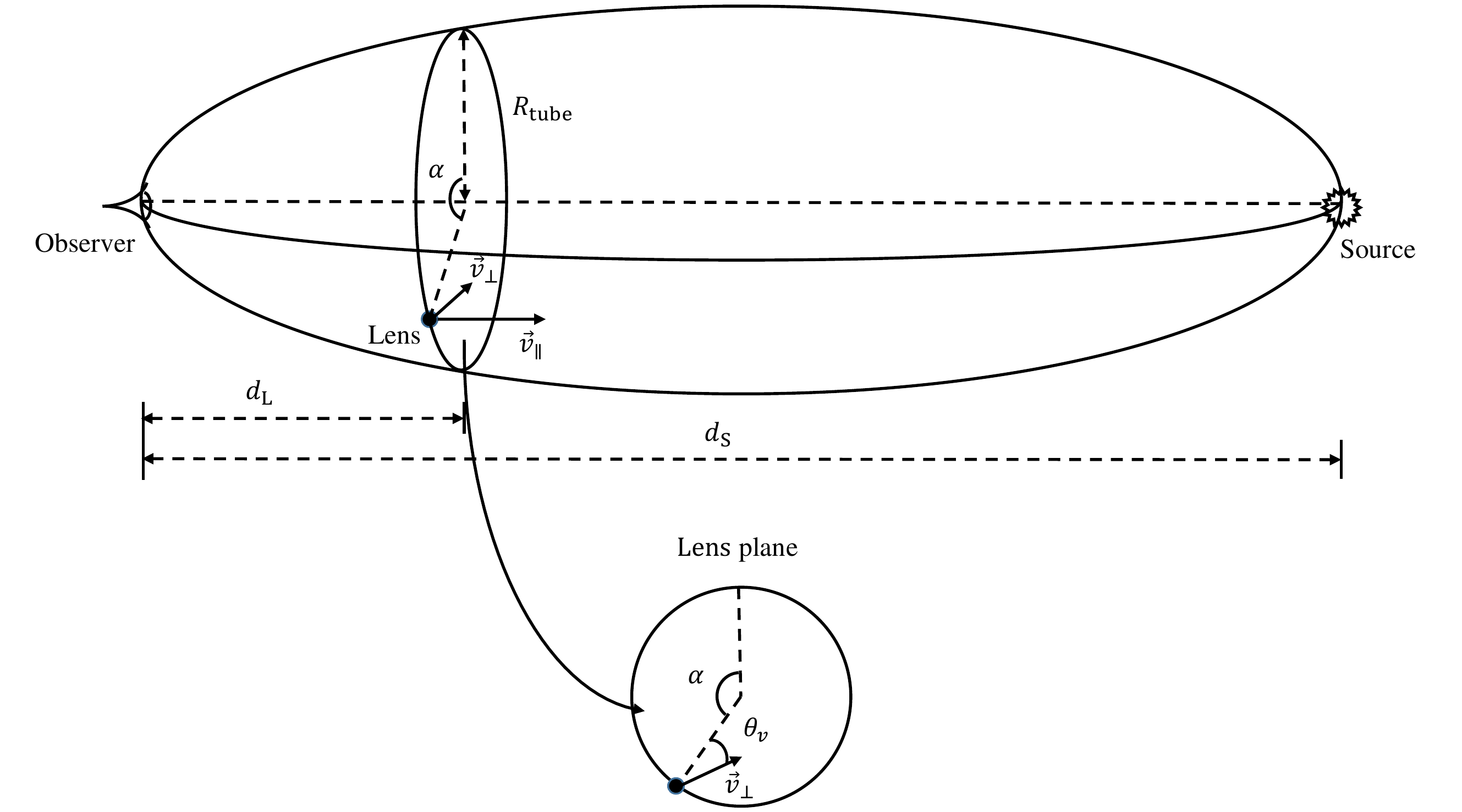}
    \caption{%
        The schematic diagram for the ``microlensing tube'' and lens plane.
    }
    \label{fig:microtube}
\end{figure}

As the lens approaches and leaves the line-of-sight direction, the magnification of the source will increase and decrease, which produces the microlensing light curve.
To compare with the observation data, one needs to calculate the differential microlensing event rate and its timescale.
Usually, the occurrence of a microlensing event is defined when the magnification can be greater than the threshold value 1.34~\cite{Niikura:2017zjd, Niikura:2019kqi}.
For the point-like lens and source, this criterion corresponds to the situation where the minimum separation between the lens and source is shorter than the Einstein radius.
For a given source, one can consider a ``microlensing tube'' with radius $\Rtube$ satisfying $\mu_{\tot}(l \leq \Rtube) \geq 1.34$ as shown in \figref{fig:microtube}, then the occurrence of a microlensing event is equivalent to a lens crossing this tube~\cite{Griest:1990vu}.
For lenses and sources with finite size, the radius of the ''microlensing tube'' $ \Rtube $ can be quite different from that of point-like cases.

The differential event rate of microlensing $ {\dif \Gamma}/{\dif \tE} $ is defined as the distribution of the frequency of microlensing events over the light curve timescale $ \tE $ for a single source per unit observational time~\cite{Niikura:2019kqi}.
The event rate can be expressed as~\cite{Griest:1990vu}
\begin{equation}
    \dif\Gamma = n_{\lens} v_\perp  \cos\theta_v \Rtube \dif \alpha \dif  \dL
    f(v_\perp,v_\parallel) v_\perp \dif \theta_v \dif v_\perp \dif v_\parallel,
\end{equation}
where $ n_{\lens} = {\rho_{\lens}}/{M_{\lens}} $ is the number density of lenses, $ \rho_{\lens} $ is the density of the lenses, $ M_{\lens} $ is the mass of the single lens, $f(v_\perp,v_\parallel) $ is the relative velocity distribution of the lens, $ v_\perp $ is the perpendicular components of the lens' relative velocity and other definitions are shown in \figref{fig:microtube}.
The timescale $ \tE $ is the lasting time for magnification larger than $ 1.34 $ or equivalently the time for the lens to cross the tube
\begin{equation}\label{}
    \tE=\frac{2\Rtube\cos\theta_v}{v_\perp}.
\end{equation}
Therefore $ {\dif \Gamma}/{\dif \tE} $ can be simplified as~\cite{Niikura:2019kqi}
\begin{equation}\label{eq:dGdtE}
    \frac{\dif \Gamma}{\dif \tE} = \pi \int_{\dL^{(\min)}}^{\dL^{(\max)}} \dif \dL \  \frac{\rho_{\lens}(\dL)}{M_{\lens}}
    \int_{-\pi/2}^{\pi/2} \dif \theta_v \ v_\perp^4 \, \tilde{f}(v_\perp),
\end{equation}
where $ v_\perp = {2\Rtube\cos\theta_v}/{\tE} $,  and $ \tilde{f}(v_\perp) $ is the distribution of the perpendicular relative velocity obtained by integrating $f(v_\perp,v_\parallel) $ on $v_\parallel$.

\section{Microlensing by dressed primordial black holes}\label{sec:microdPBH}

The accretion of dark matter around the PBHs can lead to the formation of dressed PBHs, i.e., PBHs with surrounding minihalos.
In the microlensing system, the Schwarzschild radius of PBH is relatively small and one can approximate the central PBH to be point-like.
However, the size of the surrounding minihalo can be very large and the microlensing can be greatly affected by such a surrounding minihalo.
Therefore, the microlensing by dressed PBHs can be quite different from that by bare PBHs.

\subsection{Dressed primordial black holes}

We assume that the dressed PBH is spherically symmetric.
According to the analytic and numerical results~\cite{Bertschinger:1985pd, Adamek:2019gns, Hertzberg:2019exb}, the density profile of the surrounding minihalos can be parameterized as
\begin{equation}
    \rho (r) = \begin{cases}
        \frac{3 \Mhalo}{16\pi \Rhalo^{3}} \left( \frac{r}{\Rhalo} \right)^{-9/4}, & r \leq \Rhalo, \\
        0, & r > \Rhalo,
    \end{cases}
\end{equation}
where $ \Mhalo $ is the total mass of the halo, and $ \Rhalo $ is the cutoff radius of the halo density profile.

The relation between the surrounding halo mass and central PBH mass is not quite clear, and different results are given when different assumptions are adopted.
Assuming the absence of tidal forces over PBHs, negligible  peculiar velocities, and an initial dark matter background in the Hubble flow, the halo mass is estimated to be $\Mhalo \sim (100 - 300) \MPBH$, and when the presence of the cosmological constant is taken into account, the maximum halo mass is estimated to be $\Mhalo \sim 1500 \MPBH$~\cite{Mack:2006gz, Ricotti:2007au}.
Assuming that the surrounding halo stops the accretion in the nonlinear regime when density perturbations around dressed PBHs are the order of the halo mass, the halo mass is estimated to be $\Mhalo \sim (10^{1.5} - 10^{2.5})\MPBH$ for the mass range $\MPBH \sim (10^{-8} - 10^{2})\Msun$~\cite{Berezinsky:2013fxa}.
To study the gravitational microlensing by dressed PBHs, we set $\Mhalo=100\MPBH$ in this work for simplicity and mainly focus on the impacts due to different halo radii $\Rhalo$.
Such a setting is quite reasonable, and the analysis for different halo-PBH mass relations is similar and can be easily obtained.

\subsection{Deflection angle}

To consider the microlensing by dressed PBHs, one needs to consider the combination of the central PBH and surrounding halo as the lens.
The surface mass density of such a lens can be written as
\begin{equation}
    \Sigma (\bm{b}) =\begin{cases}
        2 \int_{0}^{\sqrt{\Rhalo^2-b^2}} \rho \left(\sqrt{b^2+z^2} \right) \dif z , & b \leq \Rhalo, \\
        0 , & b > \Rhalo,
    \end{cases}
\end{equation}
where $ z $ is the coordinate along the line-of-sight direction.
Then the mass within radius $b$ in eq.~\eqref{eq:deflection} can be written as
\begin{equation}\label{eq:effmass}
    M (b) =\MPBH + 2 \pi \int_{0}^{b} \Sigma (b') b' \dif b',
\end{equation}
and the deflection angle can be expressed as
\begin{equation}\label{eq:resdefangle}
    \Dphi (b) =\frac{4 \MPBH}{b} + \frac{4 \Mhalo}{b} f(\frac{b}{\Rhalo}),
\end{equation}
with
\begin{equation}
    f(x) = 4\pi \int_{0}^{\min(x,1)} \dif x' x' \int_{0}^{\sqrt{1-x'^2}} \dif z' \tilde{\rho}(\sqrt{x'^{2}+z'^{2}}),
\end{equation}
where $\tilde{\rho}(x)$ is the reduced density profile of the surrounding minihalos, which is defined by $\rho(r) \equiv \frac{\Mhalo}{\Rhalo^{3}}\tilde{\rho}(\frac{r}{\Rhalo})$.

According to eq.~\eqref{eq:resdefangle}, one can find that the deflection angle for a dressed PBH can be determined by $ b / \Rhalo $, $ \Rhalo / \RSchPBH $ and $ \Mhalo / \MPBH $, where $\RSchPBH$ is the Schwarzschild radius of the central PBH.
It will be asymptotic to the deflection angle for a point-like lens with mass $ \MPBH $ when the distance $ b $ decreases and identical to that for a point-like lens with mass $\MdPBH \equiv \MPBH+\Mhalo$ when $b$ is larger than $\Rhalo$.
The acronym dPBH stands for ``dressed PBH''.

\begin{figure}[htbp]
    \centering
    \includegraphics[width=0.6\textwidth]{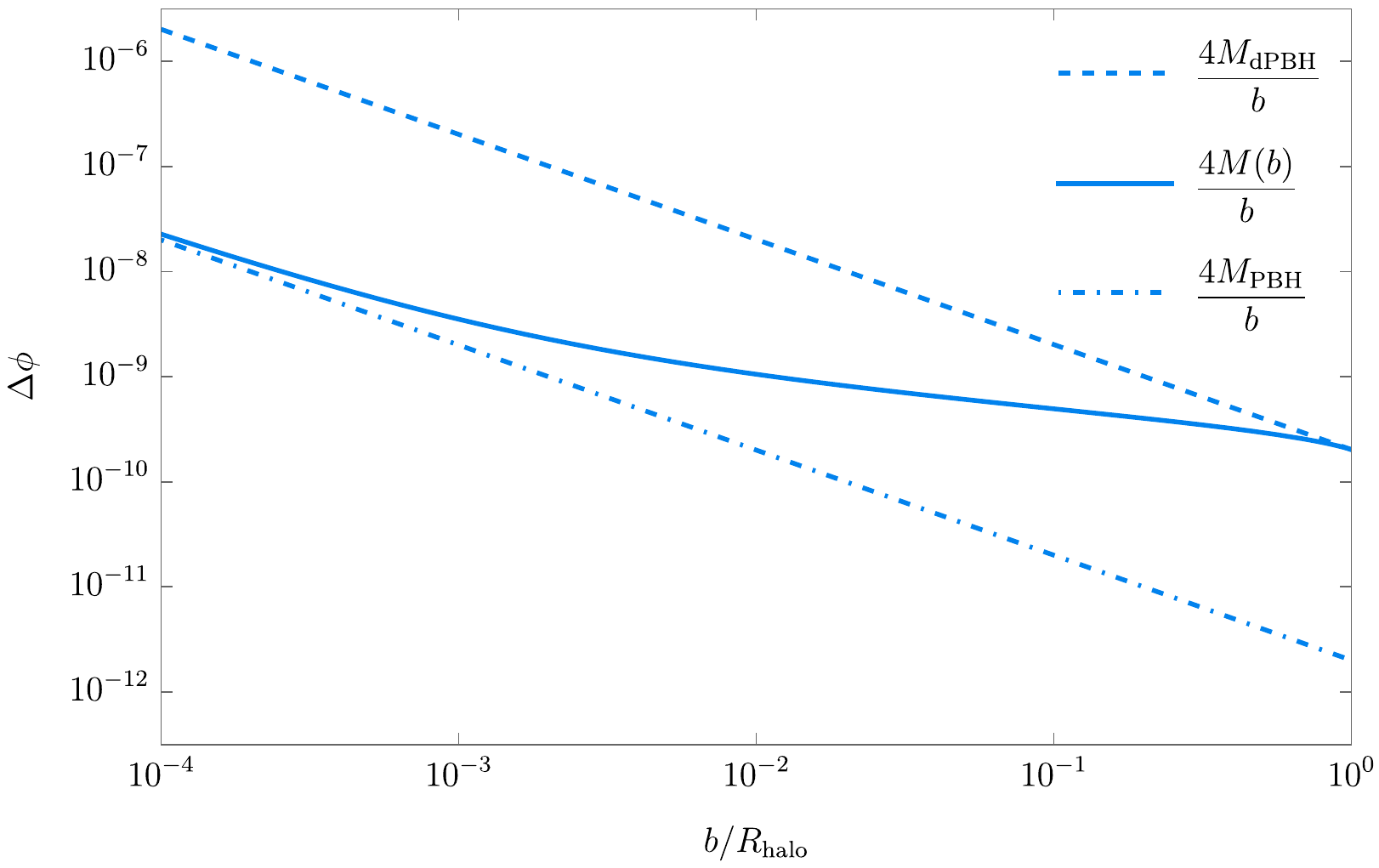}
    \caption{%
        The deflection angle for dressed PBHs with $\Rhalo=10^{12}\RSchPBH$.
        For reference, the dot-dashed and dashed lines denote a point-like lens with mass $\MPBH$ and $\MdPBH$ respectively.
    }
    \label{fig:defangle}
\end{figure}

In \figref{fig:defangle}, we show the deflection angle for a typical dressed PBH with $\Rhalo=10^{12}\RSchPBH$.
For reference, the dot-dashed and dashed lines denote a point-like lens with mass $\MPBH$ and $\MdPBH$ respectively, and such two reference lines are also shown in the following figures without redundant explanations.
This figure manifestly shows the asymptotic behavior of the deflection angle for dressed PBHs.
Although specific parameters are used in this figure, these features are universal and will help us analyze the microlensing effect produced by dressed PBHs.

\subsection{Microlensing for point-like sources}\label{sec:microlenspoint}

With eq.~\eqref{eq:resdefangle}, the imaging equation~\eqref{eq:image} can be written as
\begin{equation}\label{eq:resimage}
    \hatb^{-1} \left[ \frac{1}{1+\frac{\Mhalo}{\MPBH}} + \frac{1}{1+\frac{\MPBH}{\Mhalo}} f(\frac{\hatb}{\hatRhalo}) \right] = \hatb \pm \hatl.
\end{equation}
Here and hereafter the hat symbol denotes the quantity divided by $\REdPBH$ such as $\hatb \equiv b/\REdPBH$, where $ \REdPBH = \sqrt{4 \MdPBH D} $.
Therefore, $\hatb$ can be determined by three parameters $\{ \hatl, \hatRhalo, \Mhalo / \MPBH \}$.
For a point-like source, there are always two images with opposite parity produced by the dressed PBHs in our case.
%According to the asymptotic behavior of deflection angles, one can predict the scale of the image positions, which will simplify the numerical calculations.

Having obtained the image positions, one can calculate the magnification by using eq.~\eqref{eq:mag}.
According to eq.~\eqref{eq:resimage}, the magnification can also be determined by three parameters $\{ \hatl, \hatRhalo, \Mhalo / \MPBH \}$.
Due to the asymptotic behavior of the deflection angle, we expect that the magnification may also have asymptotic behavior.
In the gravitational microlensing phenomenon, the Einstein radius is the characteristic length.
If the halo radius $ \Rhalo $ is much larger than the Einstein radius $ \REdPBH $ of the dressed PBHs, which means that the mass within $ \REdPBH $ is mostly constituted by the central PBH, the magnification should be like the point-like lens with mass $ \MPBH $.
If $ \Rhalo $ is relatively small compared to $ \REdPBH $ which means that the halo is completely within $ \REdPBH $, the magnification should be similar to the point-like lens with mass $ \MdPBH$.

\begin{figure}[htbp]
    \centering
    \includegraphics[width=0.6\textwidth]{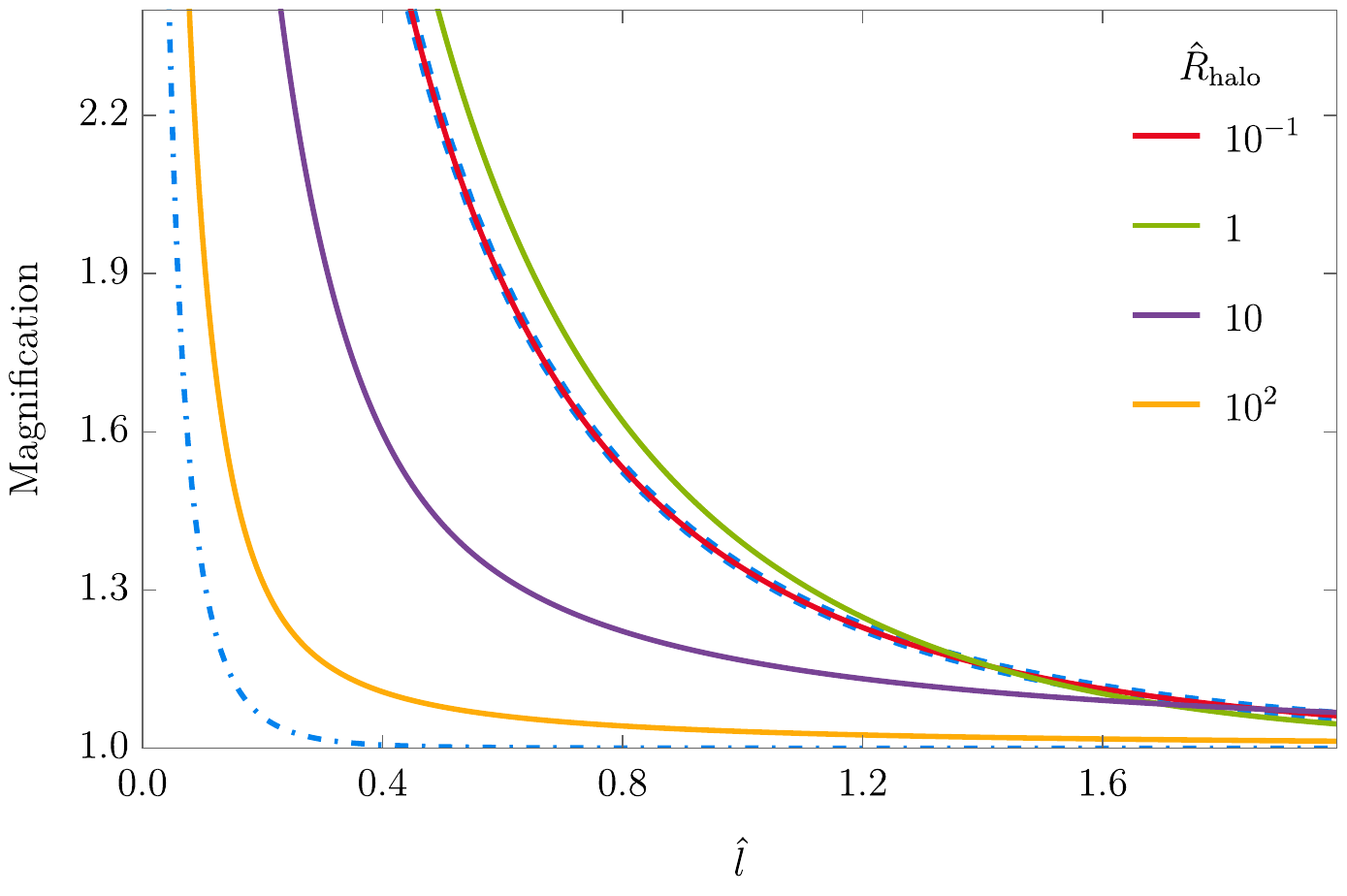}
    \caption{%
        Magnification for a point-like source from dressed PBHs with different $ \hatRhalo $.
        The red, green, purple, and yellow lines denote dressed PBHs with $\hatRhalo = 10^{-1}, 1, 10, 10^{2}$ respectively.
    }
    \label{fig:magpoint}
\end{figure}

In \figref{fig:magpoint}, we show the magnification for a point-like source from dressed PBHs with different $\hatRhalo$.
The red, green, purple, and yellow solid lines denote dressed PBHs with $\hatRhalo = 10^{-1}, 1, 10, 10^{2}$ respectively.
One can see that the yellow line (with $\Rhalo \gg \REdPBH$) approaches the dot-dashed line and the red line (with $\Rhalo \ll \REdPBH$) approaches the dashed line, which manifestly verifies our previous expectation.

\begin{figure}[htbp]
    \centering
    \includegraphics[width=0.6\textwidth]{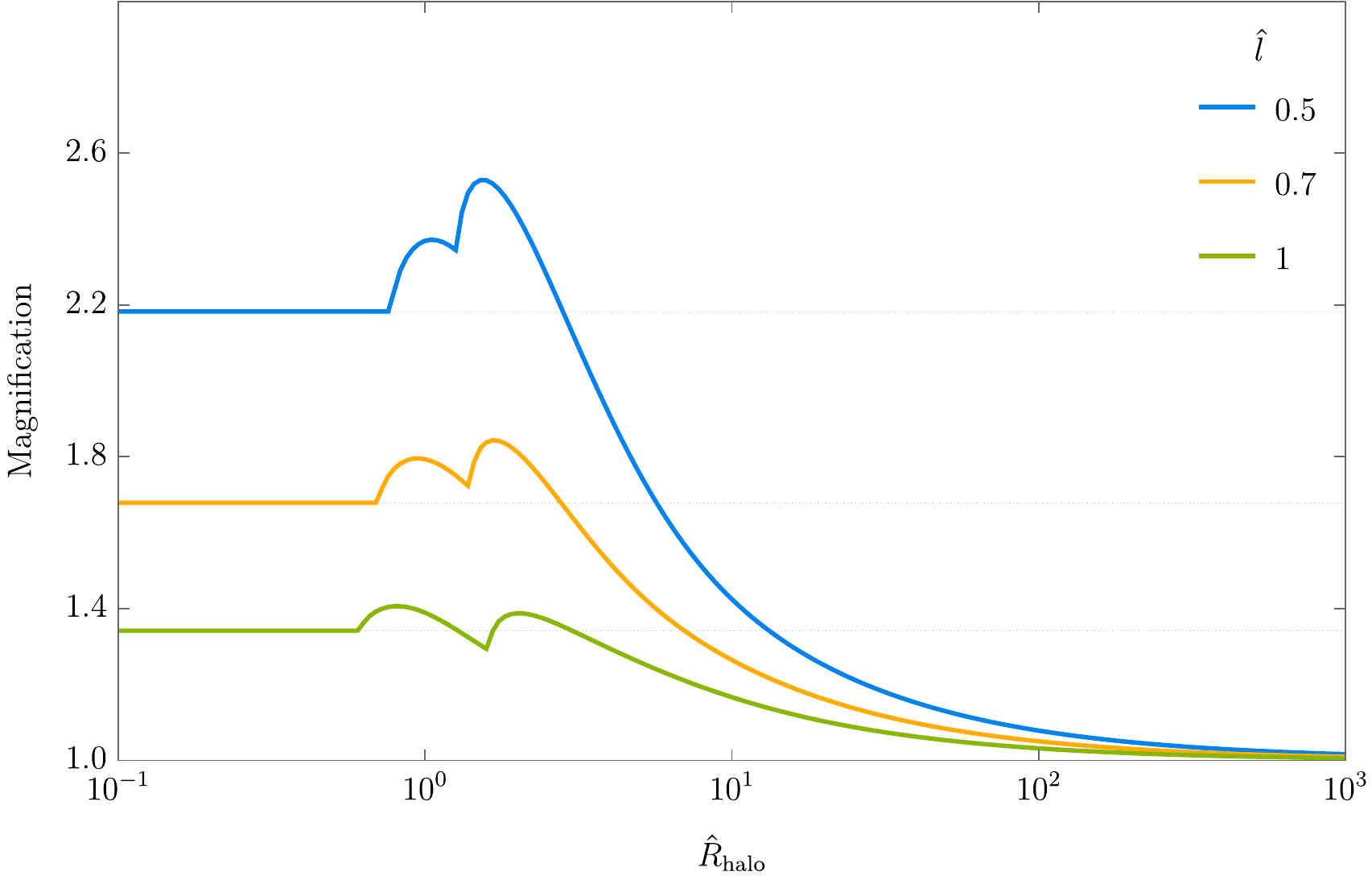}
    \caption{%
        Relation between the magnification and halo radius when the lens position is fixed.
        The blue, yellow and green lines denote lens position $\hatl = 0.5, 0.7, 1$ respectively.
    }
    \label{fig:Mag_Rhalo}
\end{figure}

One can notice that the behavior of the green line in \figref{fig:magpoint} is quite nontrivial. This is because when $\Rhalo \sim \REdPBH$ the density distribution of halo is significant and can greatly affect the gravitational lensing.
In \figref{fig:Mag_Rhalo}, we show the relation between the magnification and halo radius when the lens position is fixed.
One may expect that the magnification monotonically decreases with the halo radius increasing, but there are two regions where the magnification is enhanced.
This happens when the projected images in the lens plane are within the halo range.
The existence of a halo can make the deflection angle change more slowly and the convergence of light is enhanced.

The microlensing tube radius $ \Rtube $ is defined as a threshold distance between the lens and the line-of-sight direction, where the magnification is $ 1.34 $.
The microlensing tube radius is one of the most important parameters in microlensing and will directly influence the event rate.
According to the definition, one can directly get $ \Rtube $ from the microlensing magnification.
Therefore, the microlensing tube radius will be determined by $\{ \hatRhalo , \Mhalo / \MPBH \}$ and have similar properties as the magnification.

\begin{figure}[htbp]
    \centering
    \includegraphics[width=\textwidth]{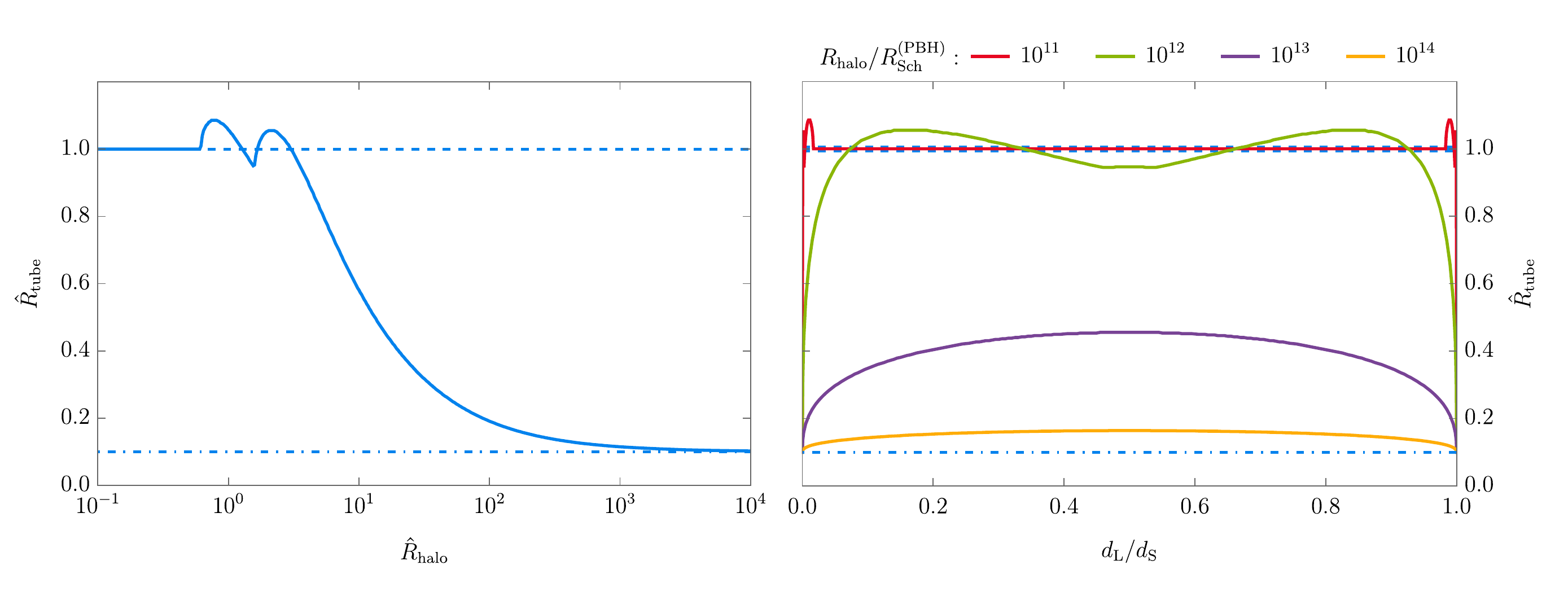}
    \caption{%
        Microlensing tube radius for a point-like source from dressed PBHs.
        The left panel shows  $\hatRtube$ with different $ \hatRhalo$.
        The right panel shows $\hatRtube$ with the same mass $ \MPBH = 10^{-5} \Msun $ and source distance $ \dS = 8\kpc$ but different $\Rhalo$ and $ \dL $.
        In the right panel, the red, green, purple, and yellow solid lines denote dressed PBHs with $\Rhalo/\RSchPBH= 10^{11}, 10^{12}, 10^{13}, 10^{14}$ respectively.
    }
    \label{fig:combinepoint}
\end{figure}

In \figref{fig:combinepoint}, we show the microlensing tube radius for a point-like source from dressed PBHs in two ways.
The left panel shows  $\hatRtube$ with different $ \hatRhalo$.
One can see that $ \hatRtube $ approaches the dot-dashed line when $\Rhalo \gg \REdPBH$ and approaches the dashed line when $\Rhalo \ll \REdPBH$.
The unusual behavior around $\Rhalo \sim \REdPBH$ is due to the enhancement of magnification as we illustrated in \figref{fig:Mag_Rhalo}.
The right panel shows $\hatRtube$ with the same mass $ \MPBH = 10^{-5} \Msun $ and source distance $ \dS = 8\kpc$ but different $\Rhalo$ and $ \dL $, which is more intuitionistic.
The red, green, purple, and yellow solid lines denote dressed PBHs with $\Rhalo/\RSchPBH= 10^{11}, 10^{12}, 10^{13}, 10^{14}$ respectively.
One can see that the yellow line (which has $\Rhalo \gg \REdPBH$) approaches the dot-dashed line and the red line (which has $\Rhalo \ll \REdPBH$) approaches the dashed line. 
The unusual shapes of the red line and green line result from the unusual behavior of $\hatRtube$ around $\Rhalo \sim \REdPBH$ as shown in the left panel.

\subsection{Microlensing for finite-size sources}\label{sec:microlensfinite}

\begin{figure}[htbp]
    \centering
    \includegraphics[width=0.6\textwidth]{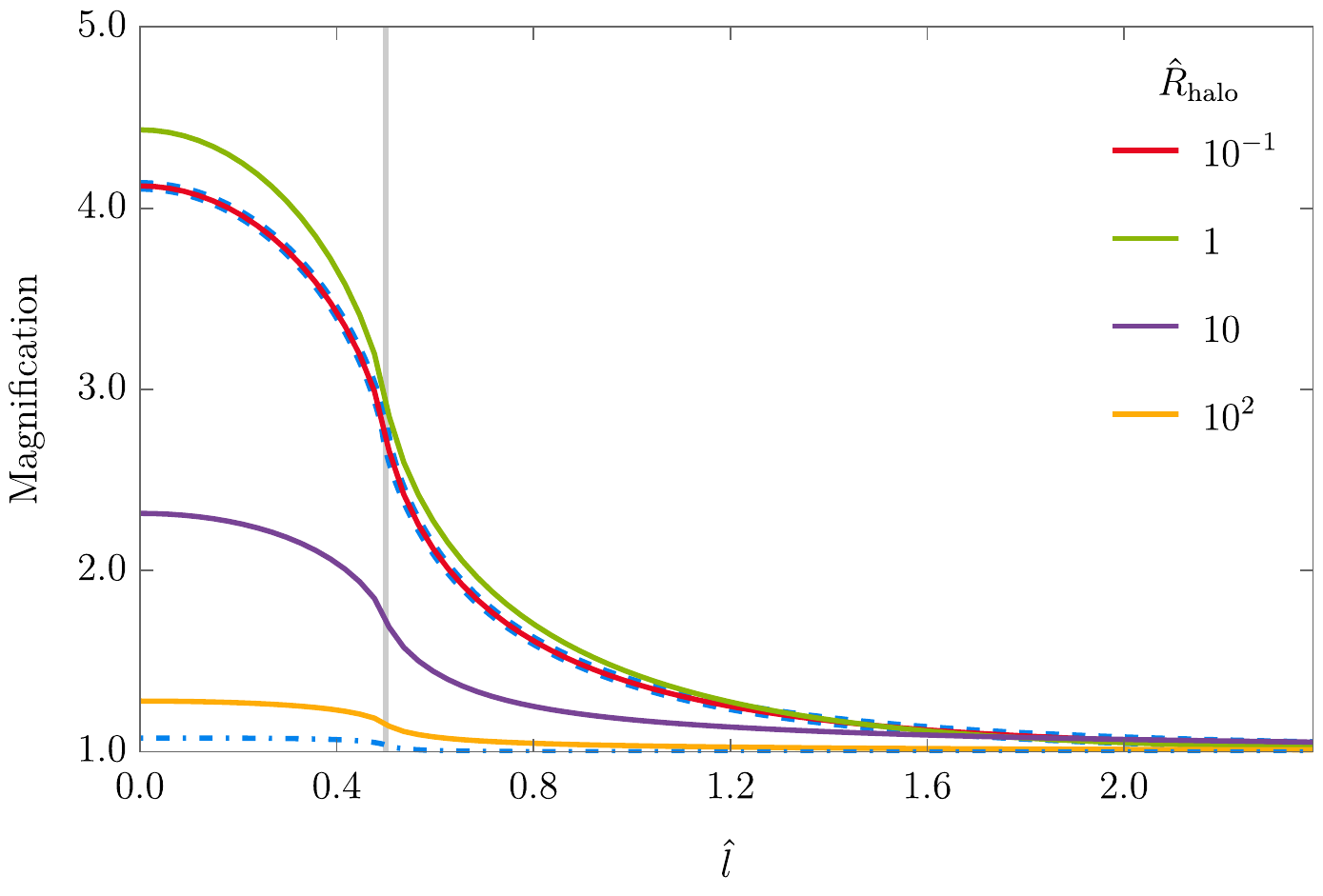}
    \caption{%
        Magnification for a finite-size source from dressed PBHs with same $ \hatrS = 1/2 $ but different $ \hatRhalo$.
        The red, green, purple, and yellow lines denote dressed PBHs with $\hatRhalo = 10^{-1}, 1, 10, 10^{2}$ respectively.
        The vertical line denotes $l =r_{\text{S}}$.
        The dashed (dot-dashed) line denotes the magnification when $\Rhalo$ is small (large) enough.
    }
    \label{fig:magfinite}
\end{figure}

Now we consider the case with finite-size sources.
Similarly, the magnification can be determined by four parameters $\{ \hatl, \hatRhalo, \hatrS, \Mhalo / \MPBH \}$, where $ \hatrS $ is an additional parameter representing the relative size of the source.
The finite source size effect is only important when the radius of the projected source is comparable with or larger than the Einstein radius.
In \figref{fig:magfinite}, we show the magnification for a finite-size source from gravitational lensing by dressed PBHs with the same $ \hatrS = 1/2 $ but different $ \hatRhalo $.
For such parameters, the Einstein radius of dressed PBH and the projected source radius are comparable, thus the finite source size effect needs to be considered.

When $ l > r_{\text{S}} $, the center of the lens is outside the projected source and the magnification for the finite-size source is similar to that for the point-like source.
When $ l < r_{\text{S}} $, the magnification for the finite-size source is totally different from the point-like source situation and has a finite maximum value for $ l=0  $.
These similarity and difference are manifestly shown by \figref{fig:magpoint} and \figref{fig:magfinite}.
The asymptotic behavior for the finite-size sources is similar to that for point-like sources and the green line representing $ \Rhalo = \REdPBH $ also has the unusual behavior.
This is because the magnification of a finite-size source can be seen as the average of magnifications produced by different pieces of it and the small pieces can be seen as point-like sources.

\begin{figure}[htbp]
    \centering
    \includegraphics[width=\textwidth]{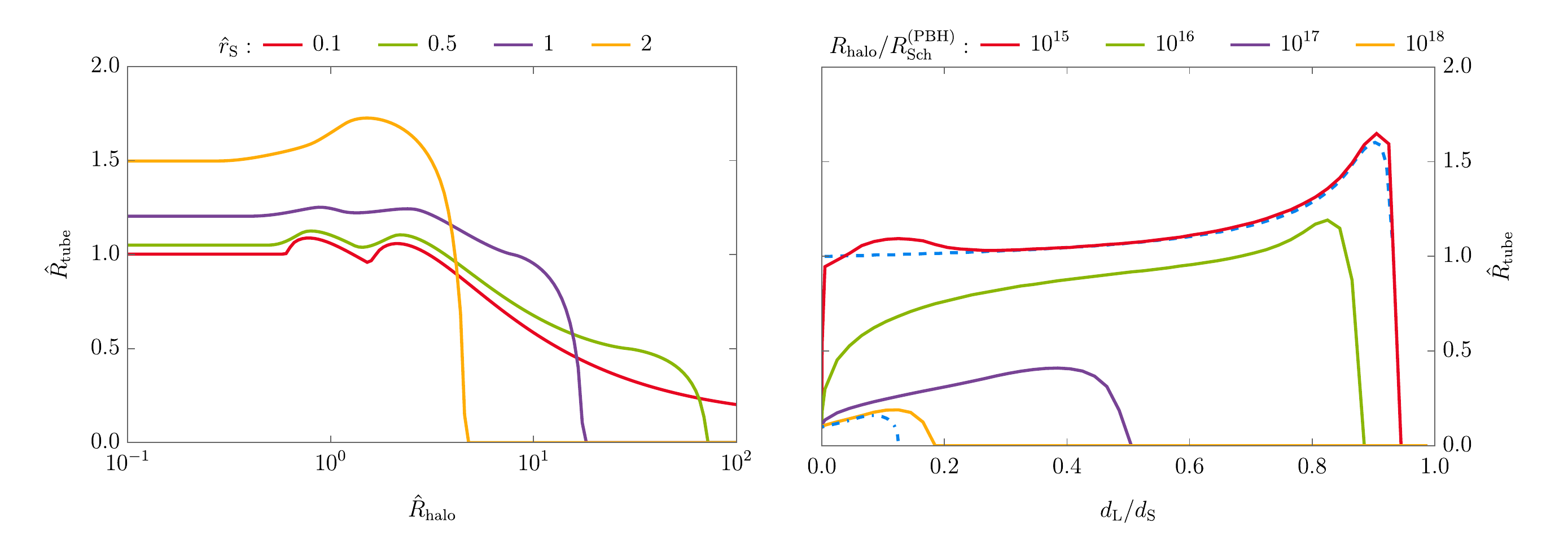}
    \caption{%
        Microlensing tube radius for a finite-size source from dressed PBHs.
        The left panel shows $\hatRtube$ with different $ \hatRhalo$ and $ \hatrS$.
        The red, green, purple, and yellow lines denote $ \hatrS = 0.1, 0.5, 1, 2$.
         The right panel shows $\hatRtube$ with the same mass $\MPBH=10^{-10}\Msun$, source radius $ R_{\text{S}} = 6.96 \times 10^8 \meter $, and source distance $\dS=770\kpc$ but different $\Rhalo$ and $\dL$.
        The  red, green, purple, and yellow lines denote dressed PBHs with $\Rhalo/\RSchPBH= 10^{15}, 10^{16}, 10^{17}, 10^{18}$ respectively.
    }
    \label{fig:combinefinite}
\end{figure}

In \figref{fig:combinefinite}, we show the microlensing tube radius for a finite-size source from gravitational lensing by dressed PBHs in two ways.
The left panel shows $\hatRtube$ with different $ \hatRhalo$ and $ \hatrS$.
The red, green, purple, and yellow lines denote $ \hatrS = 0.1, 0.5, 1, 2$.
The red line is similar to the point-like source case because the finite source size effect is weak.
For larger $ \hatrS $, the finite source size effect becomes more influential and the shape of $ \hatRtube$ changes.
With $ \hatRhalo$ increasing, the halo becomes more diffuse. 
Thus, the maximum magnification decreases and can be smaller than $ 1.34 $, which results in $ \Rtube = 0 $ for large $ \hatRhalo $.
The right panel shows $\hatRtube$ with the same mass $ \MPBH = 10^{-10} \Msun $ and source distance $\dS=770\kpc$ (which is the distance to M31), but different $\Rhalo$ and $\dL$.
The source radius is set as the solar radius ($ R_{\text{S}} = 6.96 \times 10^8 \meter $), which is the source radius set in Subaru/HSC Andromeda observation~\cite{Niikura:2017zjd}.

\section{Improved constraints on primordial black holes}\label{sec:PBHcons}

Gravitational microlensing can constrain the PBH abundance by detecting the corresponding microlensing event rate.
As shown in the last section, the minihalo surrounding PBHs can greatly affect the microlensing, therefore, the constraints on PBHs obtained through microlensing could also be affected.
In this section, we focus on the data of OGLE and Subaru/HSC Andromeda observations and give the improved constraints on PBHs from stellar microlensing.
To compare the improved constraints with original constraints, we mostly follow the analysis in~\cite{Niikura:2019kqi} for OGLE and~\cite{Niikura:2017zjd} for Subaru/HSC Andromeda observations, except that the surrounding minihalo of PBH is considered.

Assuming that the dark matter density is $ \rho_{\DM} $ and the fraction of dressed PBHs in dark matter is $ \fdPBH $, then according to eq.~\eqref{eq:dGdtE}, the expected number of microlensing events during a timescale interval of $ [\tE^{(\min)}, \tE^{(\max)}] $ can be written as
\begin{equation}\label{eq:eventnumber}
    N_{\text{exp}}^{\text{(dPBH)}} = \pi t_{\obs} N_{\text{S}} \int_{\tE^{(\min)}}^{\tE^{(\max)}} \dif\tE \, \epsilon(\tE) \fdPBH  \int_{\dL^{(\min)}}^{\dL^{(\max)}} \dif \dL \frac{\rho_{\DM}(\dL)}{\MdPBH} \int_{-\pi/2}^{\pi/2} \dif \theta_v \, v_\perp^4 \, f(v_\perp),
\end{equation}
where $ \rho_{\lens} = \rho_{\DM} \fdPBH $ and $M_{\lens}=\MdPBH$ have been used, $ t_{\obs} $ is the total observation time, $ N_{\text{S}} $ is the total number of sources, and $\epsilon(\tE) $ is the detection efficiency estimated from the simulation for the probability of a microlensing event of timescale $ \tE $ to be detected.

\subsection{OGLE observation}

The OGLE observations focus on the stars in the Galactic bulge.
The sources are assumed at the Galactic center and the spatial distribution of dark matter between the source and the observer is taken as the Navarro-Frenk-White (NFW) profile~\cite{Navarro:1996gj}
\begin{equation}
    \rho_{\DM}(r) = \frac{\rho_c}{(r/r_*)(1+r/r_*)^2},
\end{equation}
with $\rho_c = 4.88 \times 10^6 M_\odot /\kpc^3 $ and $ r_* = 21.5\kpc$.
The velocity distribution is assumed to follow the Gaussian distribution
\begin{equation}
    f(v_\perp) = \frac{1}{2 \pi \sigma^2_{\PBH}} \exp \left[-\frac{v_\perp ^2}{2 \sigma_{\PBH}^2} \right],
\end{equation}
with the square of velocity dispersion $ \sigma_{\PBH}^2 = \left[ (220)^2 +\left( \frac{\dL}{\dS} 100 \right)^2 \right] (\mathrm{km}/\mathrm{s})^2 $.
For the OGLE data, $ \dS = 8 $ kpc, $ t_{\obs} = 5 $  years and $ N_{\text{S}}  = 4.88 \times 10^7 $.
The detection efficiency $\epsilon(\tE) $ is taken as the average of detection efficiency data in~\cite{2017Natur.548..183M}.

Since most OGLE data can be fairly well reproduced by the stellar components, one can use the null hypothesis of PBH microlensing, i.e. assuming that there is no PBH lensing in the OGLE data.
This would give an upper bound on the PBH abundance.
Assuming that the microlensing events at each timescale bin follow the Poisson distribution, the likelihood of obtaining OGLE observation data is
\begin{equation}
    \mathcal{L} (\bm{N}_{\obs}  | \bm{\theta}) = \prod_{i=1}^{n_{\text{bin}}}  \frac{\lambda(\tEi)^{N_{\obs}(\tEi)} \ee^{-\lambda(\tEi)}}{N_{\obs} (\tEi)!},
\end{equation}
where $\bm{\theta}$ is the model parameter vector, $\bm{N}_{\obs}= \{ N_{\obs}(t_{\text{E},1}) , N_{\obs}(t_{\text{E},2}) , \cdots , N_{\obs}(t_{\text{E},n_{\text{bin}}}) \}$ is the data vector, $ N_{\obs}(\tEi) $ is the observed event number at the $i$-th timescale bin $ \tEi $, $ n_{\text{bin}} $ is the number of timescale bins, and $ \lambda(\tEi) $ is the expected event number at the $i$-th bin
\begin{equation}
    \lambda(\tEi) = N_{\obs}(\tEi) + N_{\text{exp}}^{\text{(dPBH)}}(\tEi),
\end{equation}
with $N_{\text{exp}}^{\text{(dPBH)}}(\tEi)$ calculated by eq.~\eqref{eq:eventnumber}.
Consider $ \fdPBH $ as a single model parameter for an assumed dressed PBHs mass scale, the distribution of $ \fdPBH $ can be computed by Bayes' theorem and one can calculate the 95\% CL upper bound on the PBH abundance $\fPBH$ by using the OGLE observation data.

\begin{figure}[htbp]
    \centering
    \includegraphics[width=0.6\textwidth]{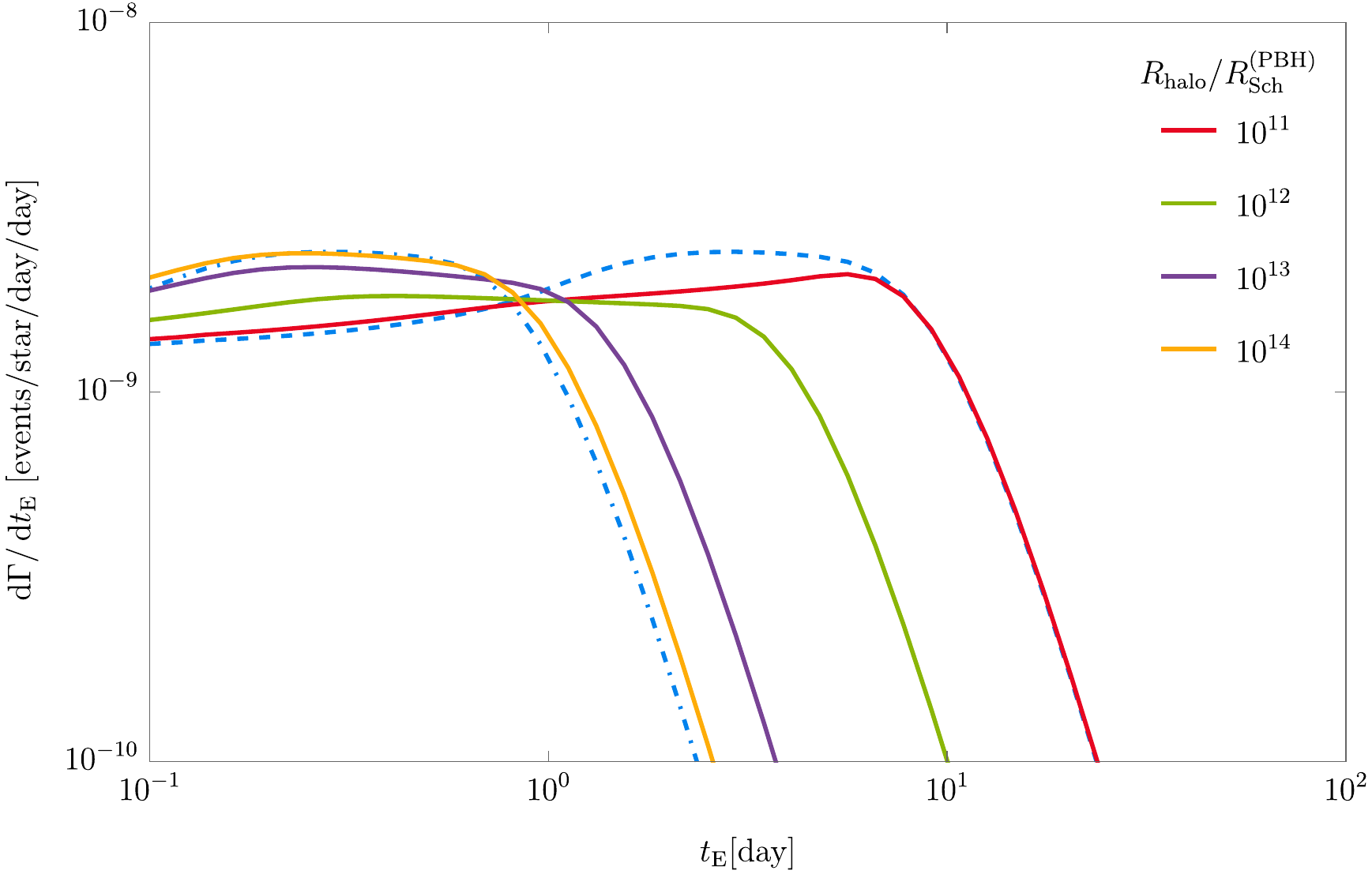}
    \caption{
        The expected differential event rate of microlensing for a single source in the MW bulge.
            The model parameters are set the same as the OGLE.
            The red, green, purple, and yellow solid lines denote dressed PBHs with $ \Rhalo / \RSchPBH = 10^{11}, 10^{12}, 10^{13}, 10^{14}$ respectively.
        For reference, the dashed (dot-dashed) line denotes the differential event rate when $\Rhalo$ is small (large) enough.
    }
    \label{fig:eventrateogle}
\end{figure}

In \figref{fig:eventrateogle}, we show the expected differential event rate in MW for dressed PBHs with the same mass $ \MPBH = 10^{-3} \Msun $ but different $ \Rhalo $ by using eq.~\eqref{eq:dGdtE}, under the assumption that all DM is made of dressed PBHs.
%    Here we assume that all DM is made of dressed PBHs with the same mass.
 %   We calculate the differential event rate per unit observation time (day), per a single source, and per unit timescale of $ \tE $ (day) using eq.~\eqref{eq:dGdtE}.
    The model parameters, such as the mass density profile and the velocity distribution, are set the same as the OGLE.
    The red, green, purple, and yellow solid lines denote dressed PBHs with $ \Rhalo / \RSchPBH = 10^{11}, 10^{12}, 10^{13}, 10^{14}$ respectively.
    For reference, the dashed and dot-dashed lines denote the differential event rate for small enough and large enough $\Rhalo$ respectively, which have the same shape according to eq.~\eqref{eq:dGdtE} and the analysis of $ \Rtube $ in section~\ref{sec:microlenspoint}.

\begin{figure}[htbp]
    \centering
    \includegraphics[width=\textwidth]{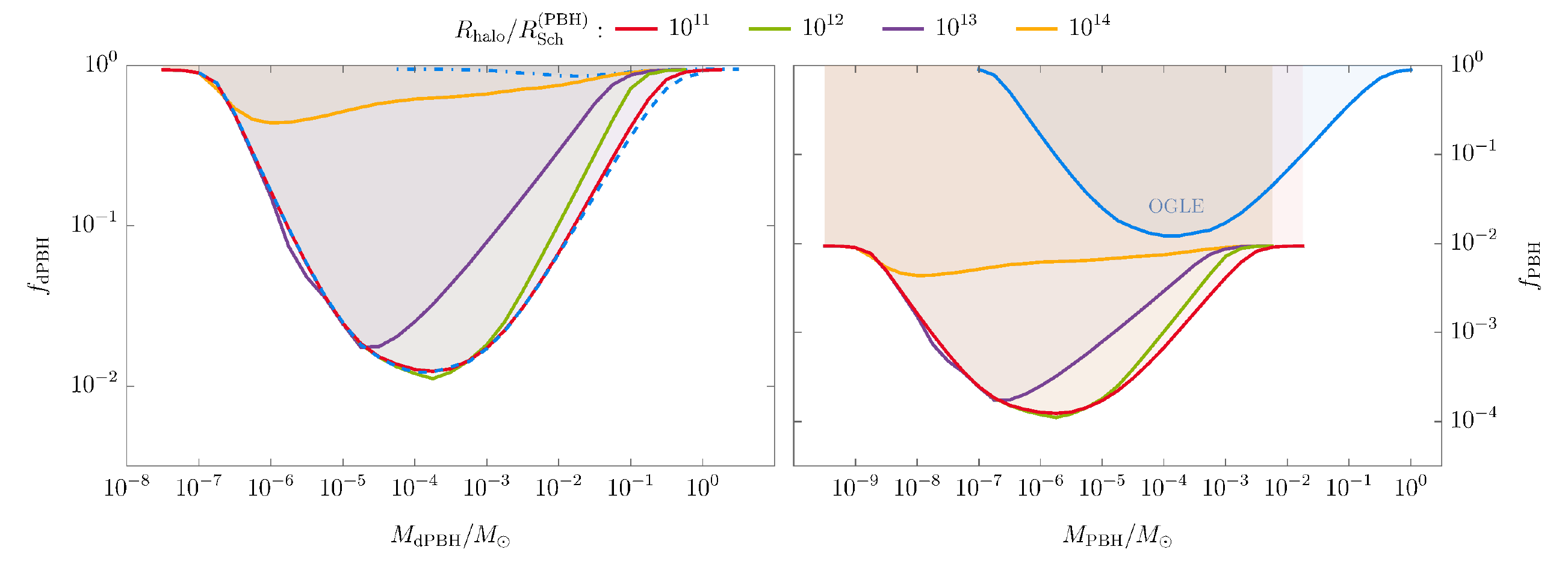}\\
    \includegraphics[width=\textwidth]{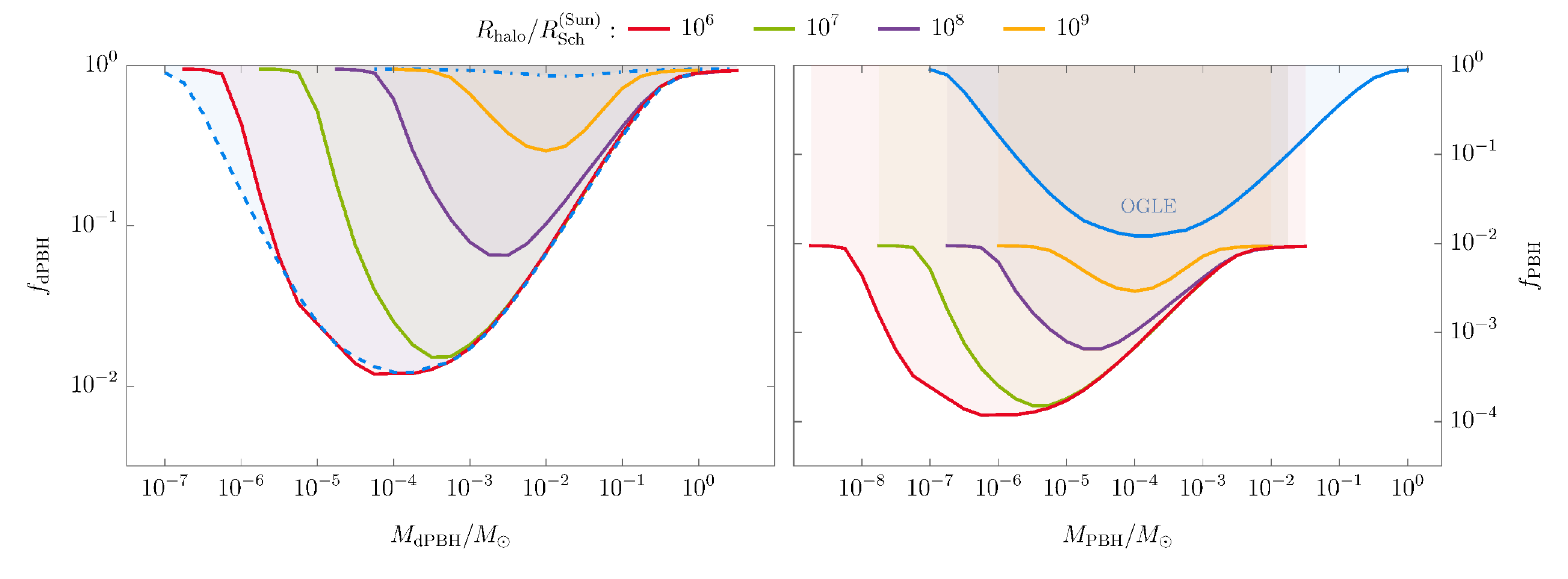}
    \caption{%
        Constraints on dressed PBHs and improved constraints on PBH from OGLE observation data.
        The upper panel shows the case that $\Rhalo \propto \MPBH$. The red, green, purple, and yellow lines denote $\Rhalo/\RSchPBH=10^{11}, 10^{12}, 10^{13}, 10^{14}$ respectively.
        The lower panel shows the case that $\Rhalo$ is independent of $\MPBH$. The red, green, purple, and yellow lines denote $\Rhalo/\RSchSun=10^{6}, 10^{7}, 10^{8}, 10^{9}$ respectively.
        The dashed (dot-dashed) line in the left panel denotes the constraints when $\Rhalo$ is small (large) enough, and the blue line in the right panel denotes the original constraint on PBH abundance without considering the surrounding minihalos.
    }
    \label{fig:combineogle}
\end{figure}

In \figref{fig:combineogle}, we show the constraints on dressed PBHs and improved constraints on PBH from OGLE observation data.
We consider two typical cases for the relation between the halo radius and central PBH mass.
In the upper panel, we consider that $\Rhalo \propto \MPBH$ and the red, green, purple, and yellow lines denote $\Rhalo/\RSchPBH=10^{11}, 10^{12}, 10^{13}, 10^{14}$ respectively.
In the lower panel, we consider that $\Rhalo$ is independent of $\MPBH$. 
The red, green, purple, and yellow lines denote $\Rhalo/\RSchSun=10^{6}, 10^{7}, 10^{8}, 10^{9}$ respectively, where $\RSchSun$ is the Schwarzschild radius of Sun.
In the left panel, the dashed (dot-dashed) line denotes the constraints when $\Rhalo$ is small (large) enough.
In the right panel, the blue line denotes the original constraint on PBH abundance without considering the surrounding minihalos.

Recalling that $\Mhalo=100\MPBH$ is used, the improved constraints on PBH abundance $\fPBH$ are obtained by using $ \MPBH = \MdPBH/101 $ and $ \fPBH = \fdPBH/101 $.
Therefore, the constraints on $ \fPBH $ plateaus at $ \fPBH = 1/101 $.
The constraints for $ \MdPBH \sim [10^{-7},10^{0}] M_\odot $ are shifted to constraints for $ \MPBH \sim [10^{-9},10^{-2}] M_\odot $.
Due to the existence of surrounding minihalos, all the constraints will be below $ 1 \% $ and stronger than the original constraints on PBHs.
When the halo size is too large so that the halo is very dispersed, the enhancement of the constraint will be very weak.
If the halo size is small, the constraint can be intensified by two orders.
Although $\Mhalo=100\MPBH$ is specified for simplicity in this work, it does not affect the fact that the existence of dark matter minihalos surrounding PBHs can greatly change the mass scale and intensity constraints for PBHs. A more detailed relation between the mass of the central PBH and the surrounding minihalo will only slightly change our results.

\subsection{Subaru/HSC Andromeda observation}

The Subaru/HSC Andromeda observations focus on the microlensing of stars in M31 by PBHs in the halo regions of the MW and M31.
Since the halos of MW and M31 are both involved, one needs to consider their total contribution.
The DM distributions of these two halos are both taken as NFW profile with $\rho_c = 4.88 \times 10^6 M_\odot /\kpc^3 , r_* = 21.5\kpc$ for MW halo and $ \rho_c = 4.96 \times 10^6 M_\odot /\kpc^3 , r_{*} = 25 $ kpc for M31 halo.
The sources are assumed at the center of M31. For a lens at $ \dL $ between the source and the observer, the distances from it to the center of MW or M31 are
\begin{align}
    r_{\text{MW}} &= \sqrt{R_\oplus^2 - 2 R_\oplus \dL \cos\theta_1 \cos\theta_2 + \dL^2}, \\
    r_{\text{M31}} &\approx \dS-\dL,
\end{align}
where $ R_\oplus = 8.5$ kpc is the distance from the Earth to the MW center, $ (\theta_1, \theta_2) = (121.2^\circ,-21.6^\circ) $ represents the angular direction of M31, and $ \dS = 770 $ kpc is the distance from the Earth to the M31 center.
The velocity distribution is given by an isotropic Maxwellian distribution
\begin{equation}
    f(v_\perp) = \frac{1}{ \pi v_c(r)^2} \exp \left[-\frac{v_\perp ^2}{ v_c(r)^2} \right],
\end{equation}
where $v_c(r) = \sqrt{{M_{\NFW}(r)}/{r}}$, $ M_{\NFW}(r) = 4 \pi \rho_c r_* ^3 [\ln (1+ r / r_*)- r/( r + r_*)] $ is the mass within radius $ r $ and needs to be calculated individually for halos of MW or M31.
For the Subaru/HSC Andromeda observation data, $ t_{\obs} = 7 $ hours, $ N_{\text{S}}  = 8.7 \times 10^7 $, $\epsilon(\tE) $ is taken as the average of detection efficiency data in~\cite{Niikura:2017zjd}.
Since there are contributions from both MW and M31 halos, the differential event rate is the sum of them.
By requiring $N_{\text{exp}}^{\text{(dPBH)}}$ to be smaller than $ 4.74 $, which is the method used in~\cite{Niikura:2017zjd}, one can get the 95\% CL upper bound on $ \fdPBH $.

\begin{figure}[htbp]
    \centering
    \includegraphics[width=0.6\textwidth]{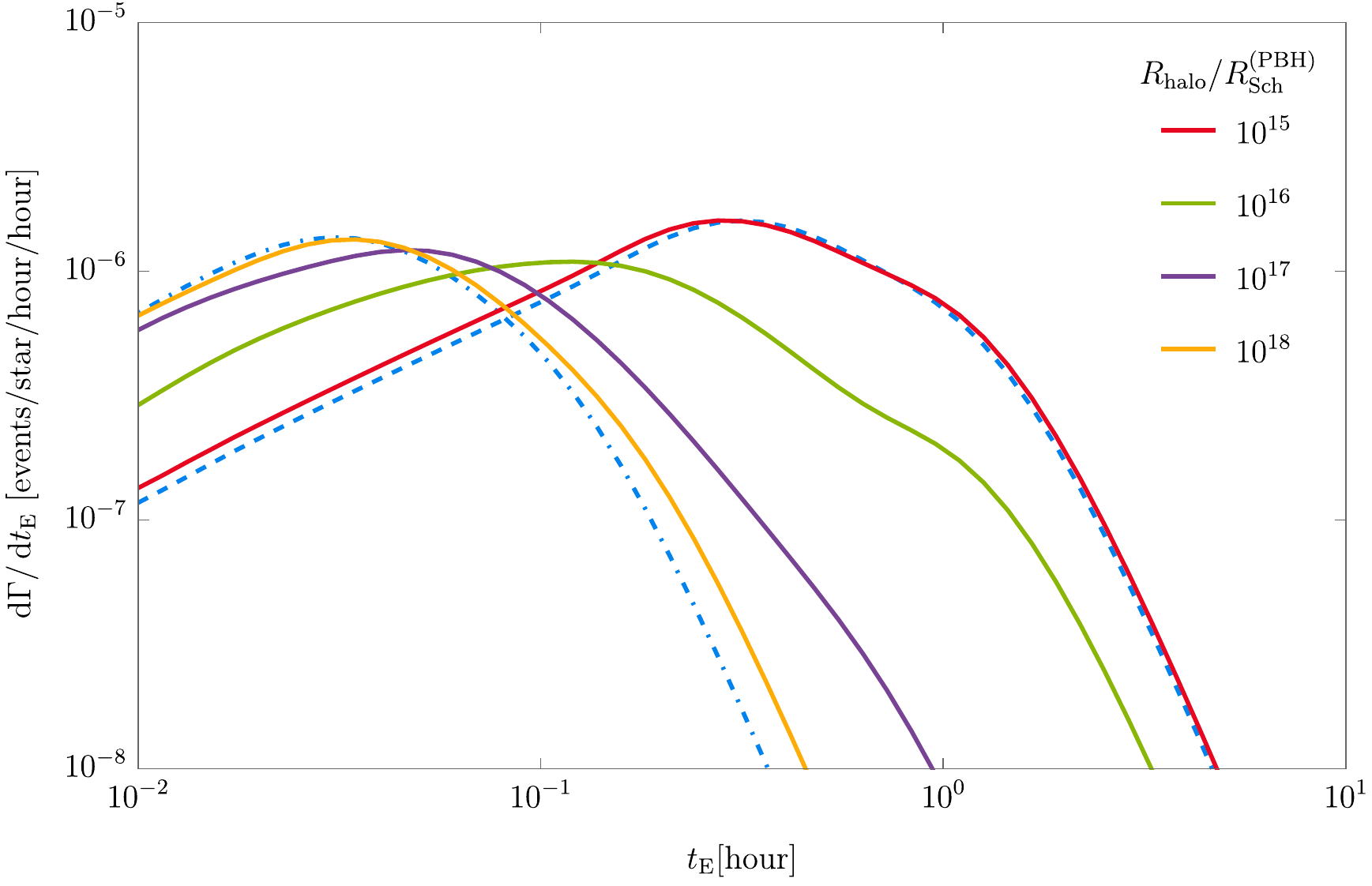}
    \caption{%
        The expected differential event rate of microlensing for a single source in M31.
        The model parameters are set the same as the Subaru/HSC Andromeda observation.
        The red, green, purple, and yellow solid lines denote dressed PBHs with $ \Rhalo / \RSchPBH = 10^{15}, 10^{16}, 10^{17}, 10^{18}$ respectively.
        For reference, the dashed (dot-dashed) line denotes the differential event rate when $\Rhalo$ is small (large) enough.
    }
    \label{fig:eventratehsc}
\end{figure}

In \figref{fig:eventratehsc}, we show the expected differential event rate in M31 for dressed PBHs with same mass $ \MPBH = 10^{-10} \Msun $ but different $ \Rhalo $, under the assumption that all DM is made of dressed PBHs.
The model parameters are set the same as the Subaru/HSC Andromeda observation.
The red, green, purple, and yellow solid lines denote dressed PBHs with $ \Rhalo / \RSchPBH = 10^{15}, 10^{16}, 10^{17}, 10^{18}$ respectively.
For reference, the dashed (dot-dashed) line denotes the differential event rate when $\Rhalo$ is small (large) enough.
In this case, the dot-dashed line and the dashed line have different shapes due to the finite source size effect.

\begin{figure}[htbp]
    \centering
    \includegraphics[width=\textwidth]{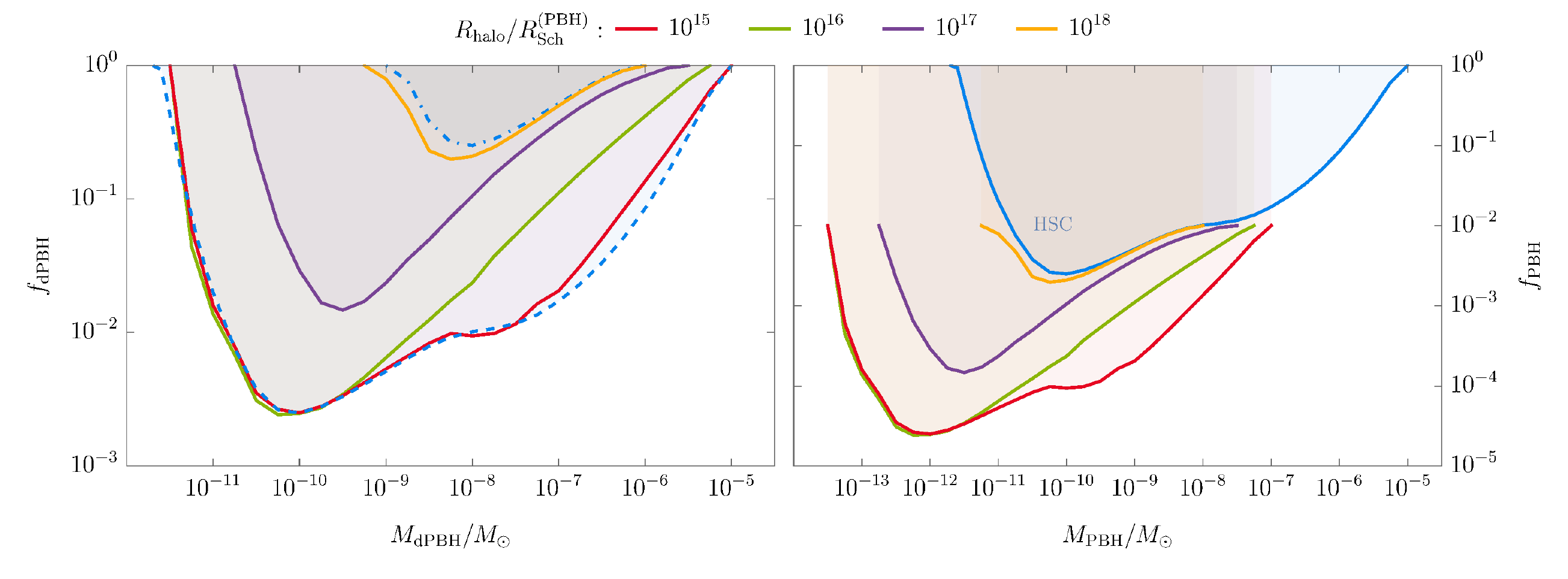}\\
    \includegraphics[width=\textwidth]{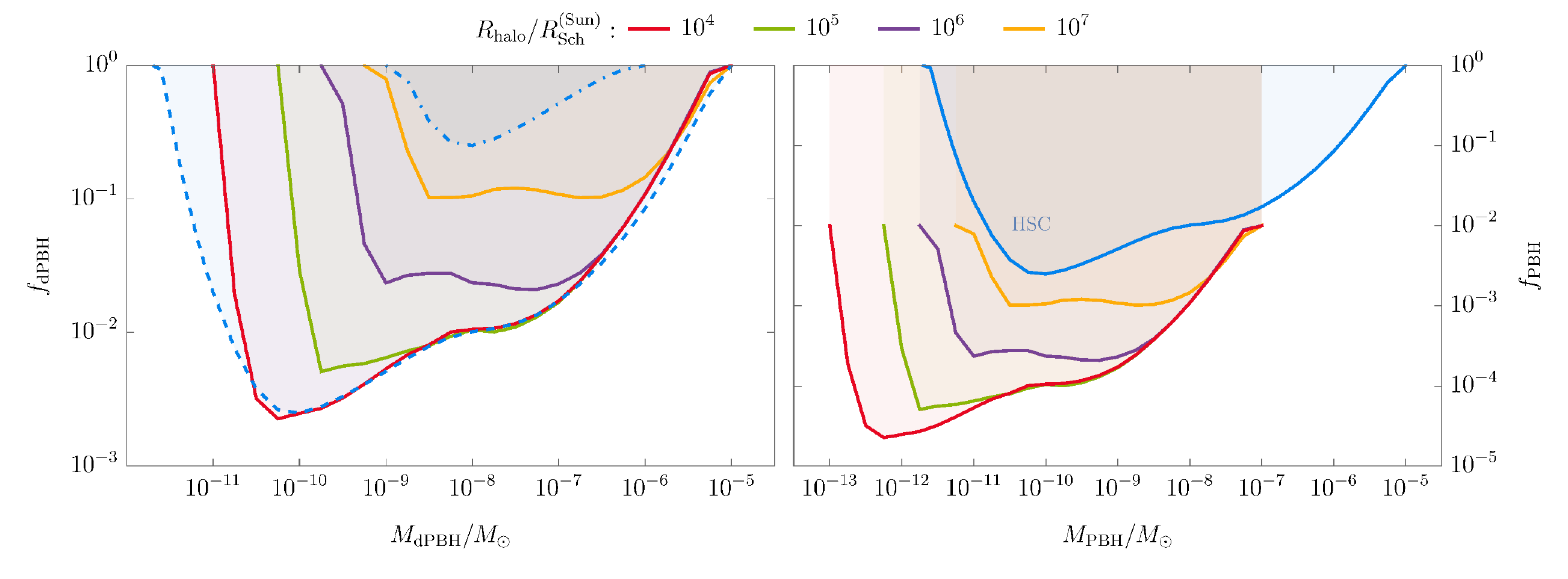}
    \caption{%
        Constraints on dressed PBHs and improved constraints on PBH from Subaru/HSC Andromeda observation data.
        The upper panel shows the case that $\Rhalo \propto \MPBH$. The red, green, purple, and yellow lines denote $\Rhalo/\RSchPBH=10^{15}, 10^{16}, 10^{17}, 10^{18}$ respectively.
        The lower panel shows the case that $\Rhalo$ is independent of $\MPBH$. The red, green, purple, and yellow lines denote $\Rhalo/\RSchSun=10^{4}, 10^{5}, 10^{6}, 10^{7}$ respectively.
        The dashed (dot-dashed) line in the left panel denotes the constraints when $\Rhalo$ is small (large) enough, and the blue line in the right panel denotes the original constraint on PBH abundance without considering the surrounding minihalos.
    }
    \label{fig:combinehsc}
\end{figure}

In \figref{fig:combinehsc}, we show the constraints on dressed PBHs and improved constraints on PBHs from Subaru/HSC Andromeda observation data, where the finite source size effect has been considered. 
In addition, we have considered both the finite source size effect and the wave optics effect for dressed PBHs with mass $ \MdPBH \leq 10^{-8} \Msun$. 
For heavier dressed PBHs, the wave optics effect can be ignored and the geometrical optics approximation is good enough. 
The meaning of denotations in \figref{fig:combinehsc} is similar to that of \figref{fig:combineogle}.
It is well known that there is a mass window $\MPBH \sim [10^{-16}, 10^{-12}]\Msun$ that PBH can constitute all the dark matter, one can find that with surrounding minihalos considered the PBHs in such a window can be constrained.

\section{Conclusion and Discussion}\label{sec:conclu}

In this work, we study the gravitational microlensing from dressed PBHs in detail.
One can see that the Einstein radius $ \RE $ is an important characteristic scale in microlensing.
For dressed PBHs, one can anticipate their microlensing effects by comparing the halo radius $ \Rhalo $ with the Einstein radius $ \RE $.
It is found that the microlensing by dressed PBHs will be asymptotic to that by the point-like lens with the mass of PBH (with the total mass of PBH and halo) if the surrounding minihalo size is much larger (much smaller) than the Einstein radius.
These asymptotic behaviors are very useful when considering the microlensing constraints on dressed PBHs.
However, this does not mean that the microlensing effects will monotonically evolve from one asymptotic limit to another.
The halo structure can greatly influence the shape of the deflection angle, images, and magnification when $ \Rhalo \sim \RE $.
In addition, there could be multiple images when a more involved minihalo profile is considered.

Applying the stellar microlensing by dressed PBHs to the data of OGLE and Subaru/HSC Andromeda observations, we obtain the improved constraints on the PBH abundance.
The constraints for dressed PBHs have similar asymptotic behavior.
The existence of dark matter minihalos surrounding PBHs can strengthen the constraints on the PBH abundance by several orders.
In addition, with the surrounding minihalos considered, the Subaru/HSC Andromeda observation data can constrain PBHs in the well-known asteroid mass window where PBHs can constitute all the dark matter.

Since the nature of dark matter is unclear, we focus on the gravitational effects which must exist and ignore other possible interactions between dark matter.
The self-interaction of specific dark matter particles like axions or WIMPs can change the halo profile and influence the constraints~\cite{Fairbairn:2017dmf, Fairbairn:2017sil, Fujikura:2021omw, Carr:2020mqm}.
Moreover, the formation and evolution of dark matter halos are very complex.
The halo profile can be influenced by the initial fraction of PBHs in dark matter, the distribution of PBHs, or the tidal force of galaxies.
Besides, the PBHs can be clustered instead of isolated.
The clustered PBHs are expected to increase accretion power and can have different halo parameters.
The disruption of halos after the formation of galaxies also needs to be considered.
Therefore, there is still a lot of work to do before we can accurately calculate the constraints for dressed PBHs and this work can be the basis for further studies.

\section*{Acknowledgments}

This work is supported in part by
the National Key Research and Development Program of China Grant  No. 2021YFC2203004, No. 2020YFC2201501 and No. 2021YFA0718304,
the National Natural Science Foundation of China Grants No. 12105344, No. 11821505, No. 11991052, No. 11947302, and No. 12047503,
the Science Research Grants from the China Manned Space Project with No. CMS-CSST-2021-B01,
and the KIAS Individual Grant QP090701.

\bibliographystyle{JHEP}
\bibliography{citeLib}

\end{document}